\documentclass{sf2a-conf2017}
\usepackage{graphicx}
\usepackage{hyperref}
\usepackage[]{natbib}  
\usepackage{epstopdf}

\def\BibTeX{{\rm B\kern-.05em{\sc i\kern-.025em b}\kern-.08em
    T\kern-.1667em\lower.7ex\hbox{E}\kern-.125emX}}
\bibpunct{(}{)}{;}{a}{}{,}  %%%%%%%%%%%%%  A&A bibliography style
\begin{document}

\TitreGlobal{SF2A 2017}

%%-----------------------------------------------------------------
%%      the top matter
%%

\title{The programme ``accurate masses for SB2 components''$^*$\footnote{* based on observations performed at the
Haute-Provence Observatory}}

\runningtitle{Accurate masses for SB2s}

\author{J.-L. Halbwachs}\address{Universit\'e de Strasbourg, CNRS, Observatoire astronomique de Strasbourg, UMR 7550, F-67000 Strasbourg, France }

\author{F. Arenou}\address{GEPI, Observatoire de Paris, PSL Research University, CNRS, Universit\'e Paris Diderot, Sorbonne Paris Cit\'e, Place Jules Janssen, F-92195 Meudon, France}

\author{H.M.J. Boffin}\address{ESO, Av. Alonso de Cordova 3107, 19001, Casilla, Santiago 19, Chile}

\author{B. Famaey$^1$}

\author{A. Jorissen}\address{FNRS \& Institut d'Astronomie et d'Astrophysique, Universit\'{e} Libre de Bruxelles, boulevard du Triomphe, 1050 Bruxelles, Belgium}

\author{F. Kiefer}\address{Institut d'Astrophysique de Paris, CNRS, UMR 7095, 98bis boulevard Arago, F-75014 Paris}

\author{J.-B. Le Bouquin}\address{UJF-Grenoble 1/CNRS-INSU, UMR 5274, Institut de Plan\'etologie et d'Astrophysique de Grenoble (IPAG), F-38041 Grenoble, France}

\author{D. Pourbaix$^4$}

\author{P. Guillout$^1$}

\author{R. Ibata$^1$}

\author{Y. Lebreton$^{7,}$}\address{LESIA, Observatoire de Paris, PSL Research University, CNRS UMR 8109, Universit\'e Pierre et Marie Curie, Universit\'e Paris}\address{Institut de Physique de Rennes, Universit\'e de Rennes 1, CNRS UMR 6251, F-35042 Rennes, France}

\author{T. Mazeh}\address{School of Physics and Astronomy, Tel Aviv University, Tel Aviv 69978, Israel}

\author{A. Nebot G\'omez-Mor\'an$^1$}

\author{L. Tal-Or}\address{Institut f\"ur Astrophysik (IAG), Friedrich-Hund-Platz 1, D-37077 G\"ottingen, Germany}

%% Keep this line, even if the page will be settled afterwards.
\setcounter{page}{237}

%%-----------------------------------------------------------------

\maketitle

%%-----------------------------------------------------------------
%%        The abstract
%% 
%%  Warning!  within the abstract:
%%  - do not use macros. 
%%  - do not use commands like: \cite, \citet, \citep ... etc.

\begin{abstract}
Accurate stellar masses are requested in order to improve our understanding of stellar interiors, but they are still rather rare. Fortunately, the forthcoming Gaia Mission will provide astrometric measurements permitting the derivation of the orbital inclinations of nearby binaries which are also observed as double-lined spectroscopic binaries (SB2s) with ground-based telescopes. A programme of radial velocity (RV) measurements was initiated in 2010 with the Sophie spectrograph of the Haute-Provence observatory in order to derive accurate SB2 orbits for a large set of stars. Therefore, combined SB2+astrometric orbits will be derived thanks to Gaia, and masses with errors around 1~\% are expected for both components. The programme includes 70 SB2s, and the accurate SB2 orbits of 24 of them were already derived. In addition, two complementary programmes
devoted to southern stars or to late-type dwarf stars were also initiated with the HERMES and the CARMENES spectrographs, respectively.
Interferometric measurements were obained with the VLTI/PIONIER for 7 SB2s, and were taken from other sources for 4 others.
Currently, combined ``visual binary'' (VB) +SB2 solutions were derived for 7 binaries, leading to the masses of the components and to the parallaxes.
The parallaxes from the Hipparcos 2 catalogue were corrected for orbital motion and compared to our solution, confirming the high quality of
Hipparcos 2.
\end{abstract}

%% Insert the keywords (to appear in the ADS indexing)
%% Keywords must be separated by a comma
\begin{keywords}
binaries: spectroscopic, binaries: visual, Astrometry
\end{keywords}

%%-----------------------------------------------------------------

\section{Introduction}
%%---------------------

Mass is the most crucial input in stellar internal structure modelling.
It predominantly influences the luminosity of a star at a given stage of its evolution, and also its lifetime.
The knowledge of the mass of stars in a non interacting binary system, together with the assumption that the components have same 
age and initial chemical composition, allows us to determine the age and the initial helium content of the system and
therefore to characterize the structure and evolutionary stage of the components.
Such modelling provides insights into the physical processes governing the structure of the stars and gives constraints 
on the free physical parameters of the models, provided the masses are known with high accuracy \citep{Lebreton05}. 
Therefore, modelling stars with extremely accurate masses (at the 1 \% level), in different ranges of masses, would 
allow to firmly anchor the models of the more loosely constrained single stars.

At present, accurate masses are still rare, but the Gaia astrometric satellite could dramatically change this situation. 
Astrometric orbits will be obtained for several systems which are already known as spectroscopic binaries (SBs). When the 
radial velocities (RVs) of both components of an SB are measured, i.e. for double-lined SBs (SB2s), the products 
${\cal M}_1 \sin^3 i$ and ${\cal M}_2 \sin^3 i$ may be derived from the orbital elements; therefore, when the inclination 
$i$ of the orbit will be derived from Gaia observations, so will be the masses of the components, ${\cal M}_1$ and ${\cal M}_2$.
In addition, the semi-major axis of a photocentric orbit is related to the luminosity ratio of the 
components, allowing to derive the individual magnitudes in the Gaia $G$ band. 
  
For all these reasons, an observational programme was initiated in 2010, using the SOPHIE spectrograph and the 193~cm telescope 
of the Observatory of Haute-Provence (OHP). This programme is presented in details in Sections~\ref{sec:targets} and \ref{sec:status}.
In addition, two complementary RV programmes were initiated on other telescopes, and also 
an interferometric programme. These related programmes are presented in Sections~\ref{sec:others} and \ref{sec:interfero} hereafter.
Thanks to the RVs and to the interferometric measurements, it was possible to derive the masses of the components of a few binaries, as well as
the parallax of these systems. The Hipparcos parallaxes were thus verified, as explained in Section~\ref{sec:parallaxes}.

\section{The target list}
%%-------------------------
\label{sec:targets}

\begin{figure}[ht!]
 \centering
 \includegraphics[width=0.8\textwidth,clip]{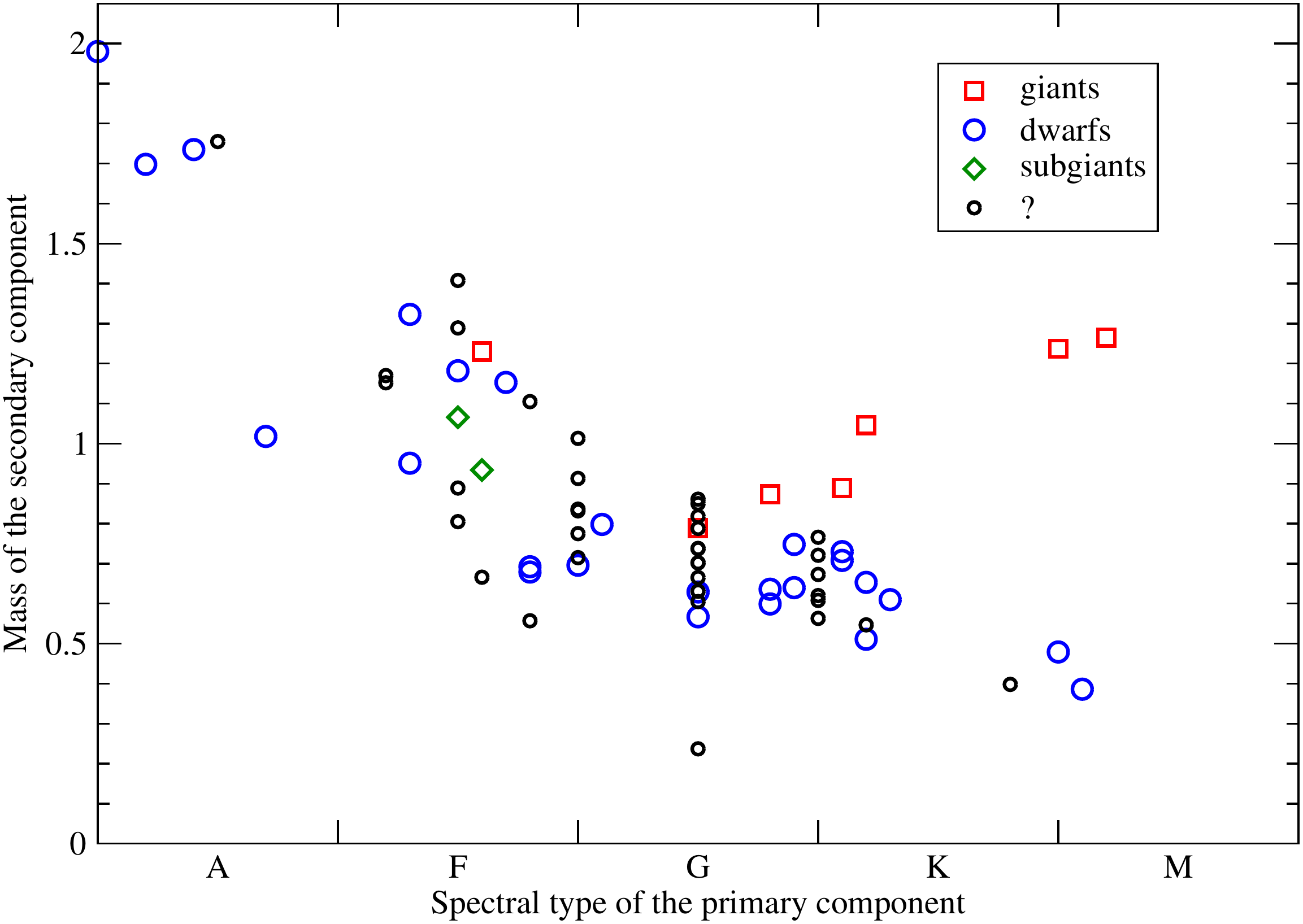}      
%% Note the ABSENCE of the extension .pdf  !
  \caption{The ``primary spectral type vs secondary
mass'' diagram of the main sample.}
  \label{Halbwachs2:fig1}
\end{figure}

Two lists of targets are used to manage the OHP programme. The principal one, which is called the ``main sample'' hereafter, was extracted from a selection of
about 200 SBs (single-lined or double-lined) fainter than 6th mag and for
which the probability to obtain the component masses with an accuracy better than 1~\% was estimated to be
larger than 20~\%, at least if it is possible to derive the RV of the secondary component. 
Seventy SB2s were eventually retained on the basis of our first observations, including 24 which were previously known as SB1s
and for which the secondary component was detected with SOPHIE. The selection process is described in details in \cite{paper1}. 
The spectral types of the primary components
are between A0 and M2. The majority of stars are on the main sequence, but 6 SB2s have a late-type giant primary component.
Rough estimates of the masses of the secondary components may be derived from the spectral types of the primary components and
from the mass ratios of the systems. Masses between 0.3 and 2~${\cal M}_\odot$ are then obtained.
The ``primary spectral type vs secondary mass'' diagram of the main sample is shown in Fig.~\ref{Halbwachs2:fig1}.

In 2013, it was decided to observe also targets brighter than 6th mag. So bright stars are observed with Gaia, but it is 
expected that their astrometric measurements will be less accurate than those of the moderately faint stars. Above all, 
these bright targets were used as backup targets when the weather was too bad to observe the main sample.
In practice, it was then possible to perform a few observations with bad weather conditions, but, fortunately, this
happened rarely and the SBs of the backup sample received very few observations. In practice, however, the backup sample
includes some SB1s (4 so far) for which we were able to detect the secondary spectrum.

\section{Present status}
%%--------------------
\label{sec:status}

\begin{figure}[ht!]
 \centering
 \includegraphics[width=0.8\textwidth,clip]{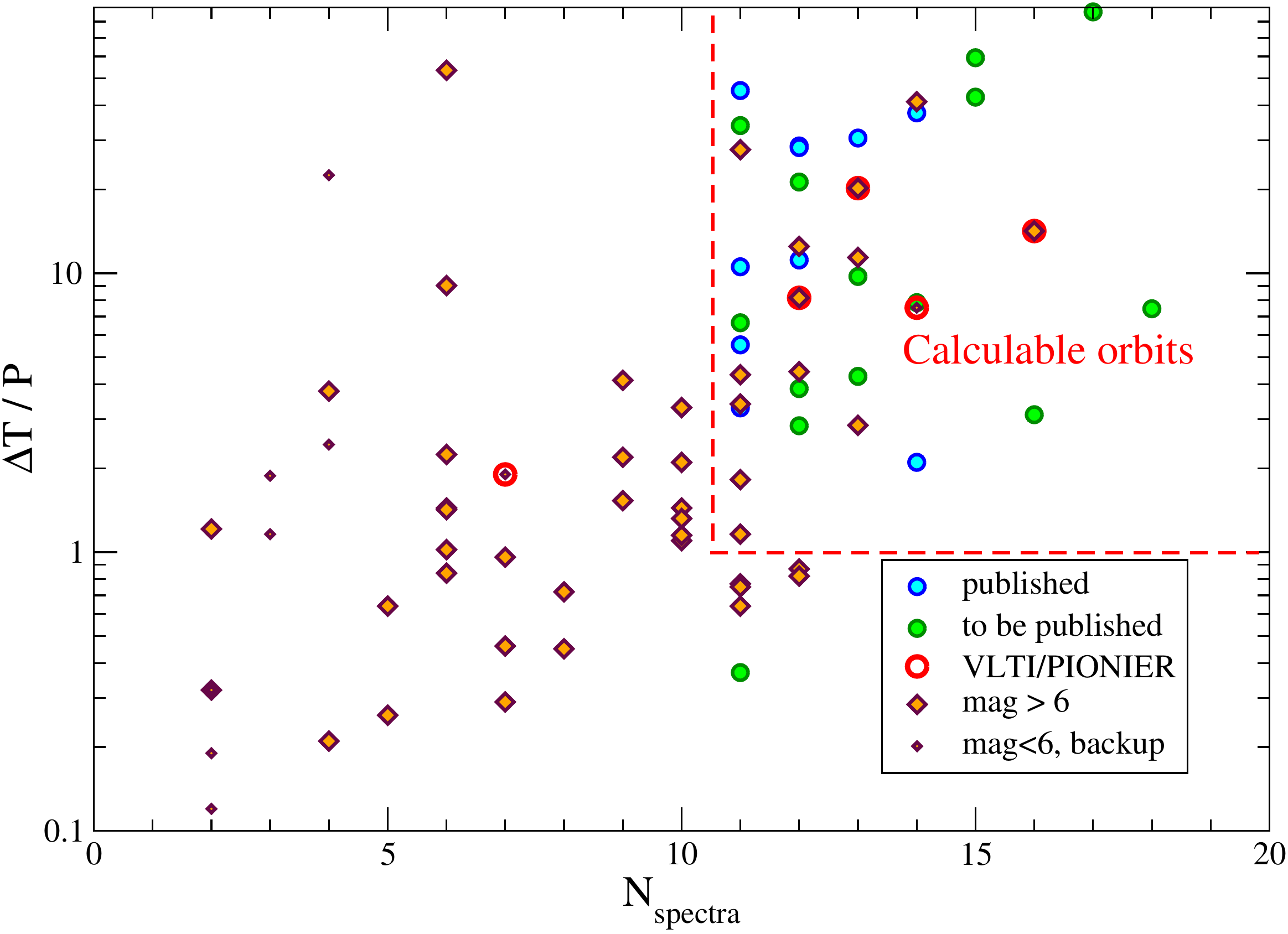}      
%% Note the ABSENCE of the extension .pdf  !
  \caption{The ``number of spectra vs orbit coverage'' diagram of the sample. The orbital elements may be
calculated for the stars in the upper right corner, when the RV of the secondary was actually derived at least 11
times, and when the measurements are distributed all over the orbit.
}
  \label{Halbwachs2:fig2}
\end{figure}

In early August 2017, we had collected 1183 spectra, and we had also found 21 additional ones in the SOPHIE archive. The status of the
main sample is presented in Fig.~\ref{Halbwachs2:fig2}, which is a ``orbit coverage vs number of spectra'' diagram.
The orbital elements of an SB2 may be reliabily derived when the number of RV measurements of each component is
larger than 11: this limit makes possible to derive SB1 orbits with at least 5 degrees of freedom,
and therefore to estimate the weights of the RVs of each component in the computation of the elements of the SB2 orbit
\citep[see][]{paper2}.
On another side, it is requested that the orbit is covered by the observations, i.e. that the time span
$\Delta T$ is longer than the period. The area delimited in the upper right corner of Fig.~\ref{Halbwachs2:fig2}
contains the SB2s for which the orbital elements could be derived. Nevertheless, additional observations are still
required for some of these stars: the secondary RV is not always measurable, and the RV measurements must also
be distributed among all the phases of the orbit.

The orbital elements of one SB2 were derived although it was not observed over a complete period: HIP 77122
has an eccentric orbit with a 11-year period, and it was observed near the periastron which is the only part of the
orbit where it is possible to derive the RVs of both components. So, we have derived an accurate orbit by fixing the period
to a value obtained taking into account old measurements.

%\section{Results}

\begin{figure}[ht!]
 \centering
 \includegraphics[width=0.48\textwidth,clip]{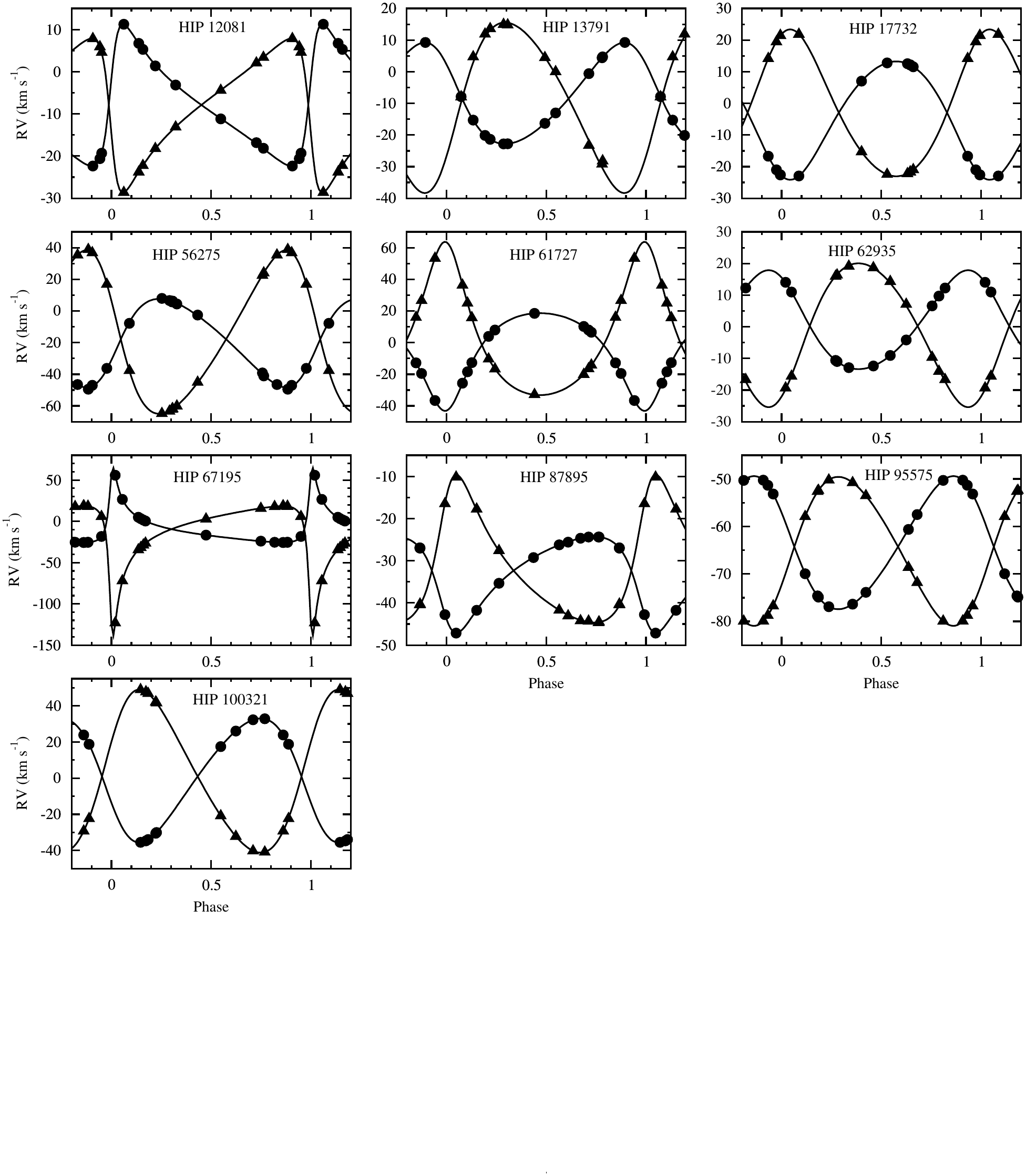}   
 \includegraphics[width=0.48\textwidth,clip]{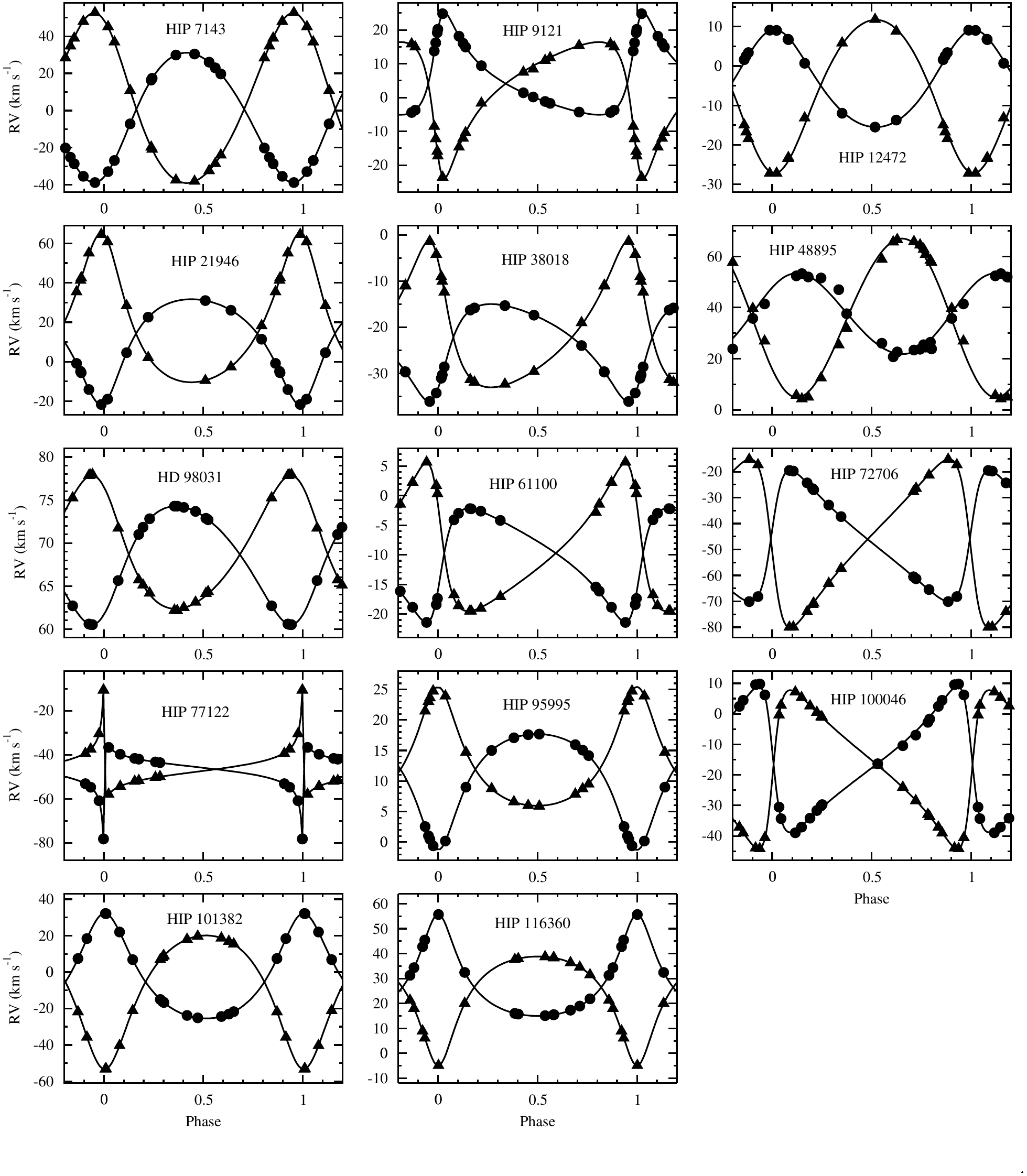} 
%% Note the ABSENCE of the extension .pdf  !
  \caption{Left: The 10 spectroscopic orbits derived from the RV measurements obtained with TODCOR in \cite{paper3}.
Right: The 14 additional spectroscopic orbits derived from the RV measurements obtained with TODCOR \citep{paper4}}
  \label{Halbwachs2:fig3}
\end{figure}

The RVs of the binary components are derived twice: a preliminary estimation is computed from the cross-correlation
function (CCF) of each spectrum with a template. This CCF is a product of the Sophie reduction pipe-line, and the
quality of the RVs thus obtained is quite sufficient to monitor the observations: these RVs may be used to improve
the orbital elements used to compute ephemerides and to plan observations, and possible outliers may be detected.
Nevertheless, the final RVs are derived from a new reduction using the TODCOR algorithm \citep{TODCOR,TODCOR2}. They are 
significantly different from the preliminary values, but more reliable, as shown in \cite{CCFvsTODCOR}.
So far, our observations lead to high-quality orbital elements for 24 SB2s: 10 SB2s in \cite{paper3}, % (Fig.~\ref{Halbwachs2:fig3}),
and 14 in \cite{paper4} (Fig.~\ref{Halbwachs2:fig3}). Among the 48 components of these binaries, 32 have minimum masses
more accurate than 1~\%.

\section{Other RV programmes}
\label{sec:others}

The OHP programme was complemented with two others, which are based on a similar
selection. These programmes are presented hereafter:

\begin{enumerate}
\item
{\bf The HERMES programme}.
Since the latitude of 
OHP is around +44~$\deg$ we had not selected SBs with declination lower than -5~$\deg$. However, the stars 
with declination between -5 and -30 $\deg$ are easily observable from the La Palma Observatory (Canary Islands) with the HERMES
spectrograph mounted on the Mercator telescope \citep{Raskin2011}. Fifty-height SB1s and SB2s were selected in a first step, 
and, after some observations, 7 of them were eventually retained.
\item
{\bf The CARMENES programme}.
About 150 SB1s were observed with SOPHIE, but the secondary component was detected for only 20 of them.
The CARMENES spectrograph mounted on the 3.5m telescope of Calar Alto is much more efficient than SOPHIE
to detect late-type secondaries, since it is working in the infrared range. Twenty-three SB1s were selected
among the OHP targets because their secondary components were expected to have spectral types at least 
as late as M. After one year, the secondary component was detected for 9 SBs, and it is even possible
to derive an accurate orbit for 2 of them. For the 7 others, the number of RV measurements is too small
to improve the orbital elements, but it is possible to derive the mass ratio.
\end{enumerate}

\section{SB2 resolved by interferometry}
\label{sec:interfero}

Gaia will provide astometric orbits, i.e. the motion of the photocentre of each binary around the barycentre,
in addition to
the single-star parameters, which are the position, the proper motion and the trigonometric parallax.
Contrarily to an astrometric orbit, an interferometric orbit is a ``visual binary'' (VB) orbit, describing the motion of one component 
(usually, the faintest one) with respect to the brightest one. However, a combined SB2+VB
orbit provides most of the elements of an SB2+astrometric orbit: it includes the period and the eccentricity,
the masses of the components, the three angles defining the orientation of the orbit, and also the
parallax. Therefore, it is highly relevant to collect enough interferometric measurements to derive
combined SB2+VB orbits for some stars of our sample. This will make possible a verification
of the masses derived for our programme, but also of the parallaxes. 

\begin{figure}[ht!]
 \centering
 \includegraphics[width=0.9\textwidth,clip]{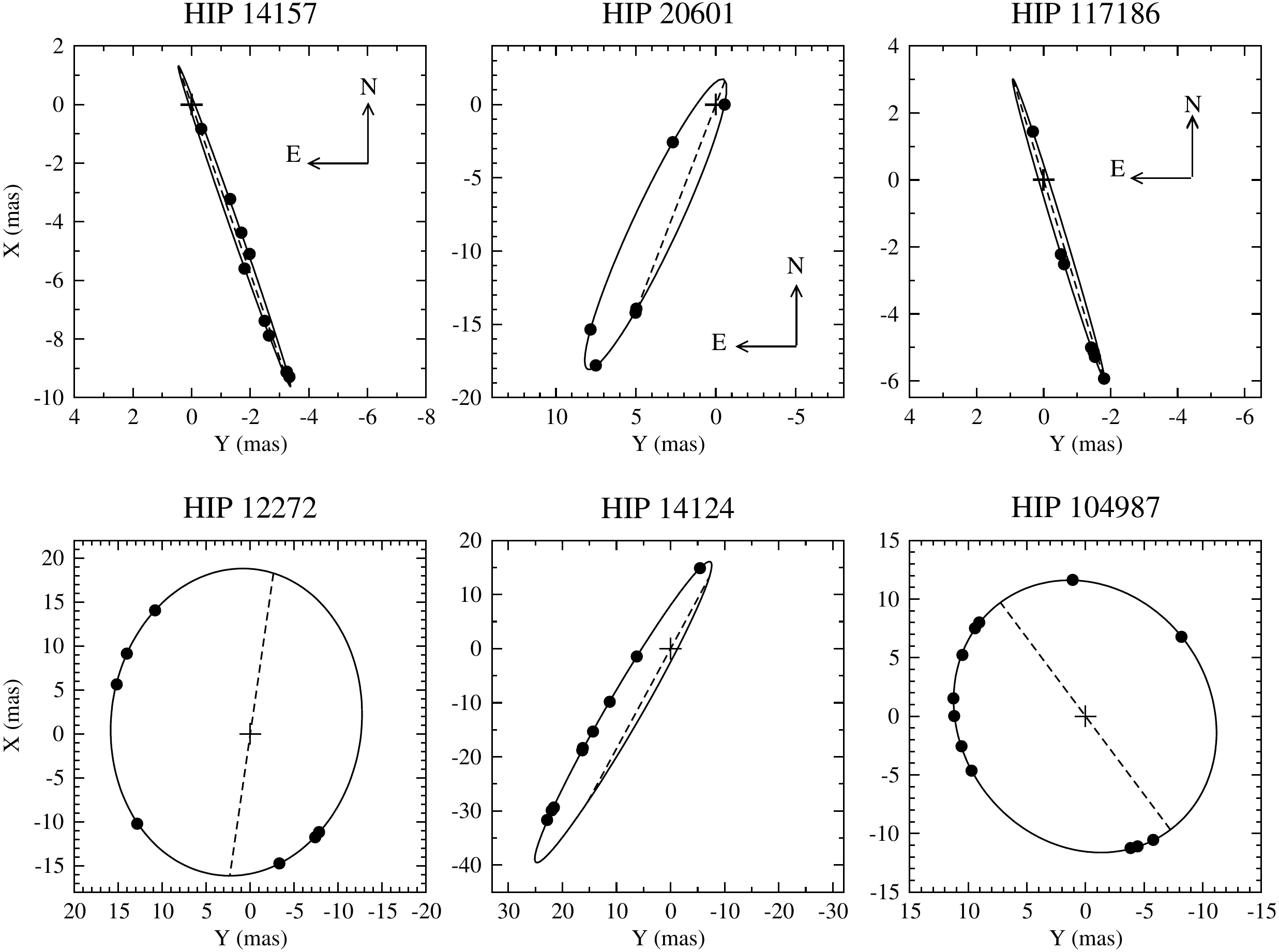}      
%% Note the ABSENCE of the extension .pdf  !
  \caption{Up : The 3 first interferometric orbits obtained with PIONIER \citep{paper2}.
Down : Three other PIONIER orbits \citep{Boffin17}.}
  \label{Halbwachs2:fig4}
\end{figure}

Among the 24 SB2s with improved orbital elements, 4 binaries were resolved by interferometry in the past, and they were sufficiently
observed to derive their orbital inclinations. In addition, we obtained interferometric observations
with
the PIONIER instrument of the ESO Very Large Telescope Interferometer (VLTI) for
7 binaries: 3 SB2s of the OHP main sample, 2 of the OHP backup sample,
and 2 of the HERMES program. The apparent relative orbits of 3 of these binaries were derived in \cite{paper2}. They are
presented in Fig.~\ref{Halbwachs2:fig4} with three others that will be published soon \citep{Boffin17}.

\section{Verification of the Hipparcos parallaxes}
\label{sec:parallaxes}

It is too early to verify the Gaia parallaxes of astrometric binaries, but we can
still verify the results of the Hipparcos mission. We consider the Hipparcos 2 reduction
\citep{hip2}, but we can't compare directly our parallaxes with those provided in the catalogue:
The Hipparcos 2 parallaxes were computed ignoring the orbital motion, and they must be corrected
on the basis of a combined SB2+astrometric solution. In practice, the correction is not important for
most of the binaries, but it is really significant when the period is close to one year \citep[see \textit{e.g.}]
[]{PourbJor00}

\begin{figure}[ht!]
 \centering
 \includegraphics[width=0.8\textwidth,clip]{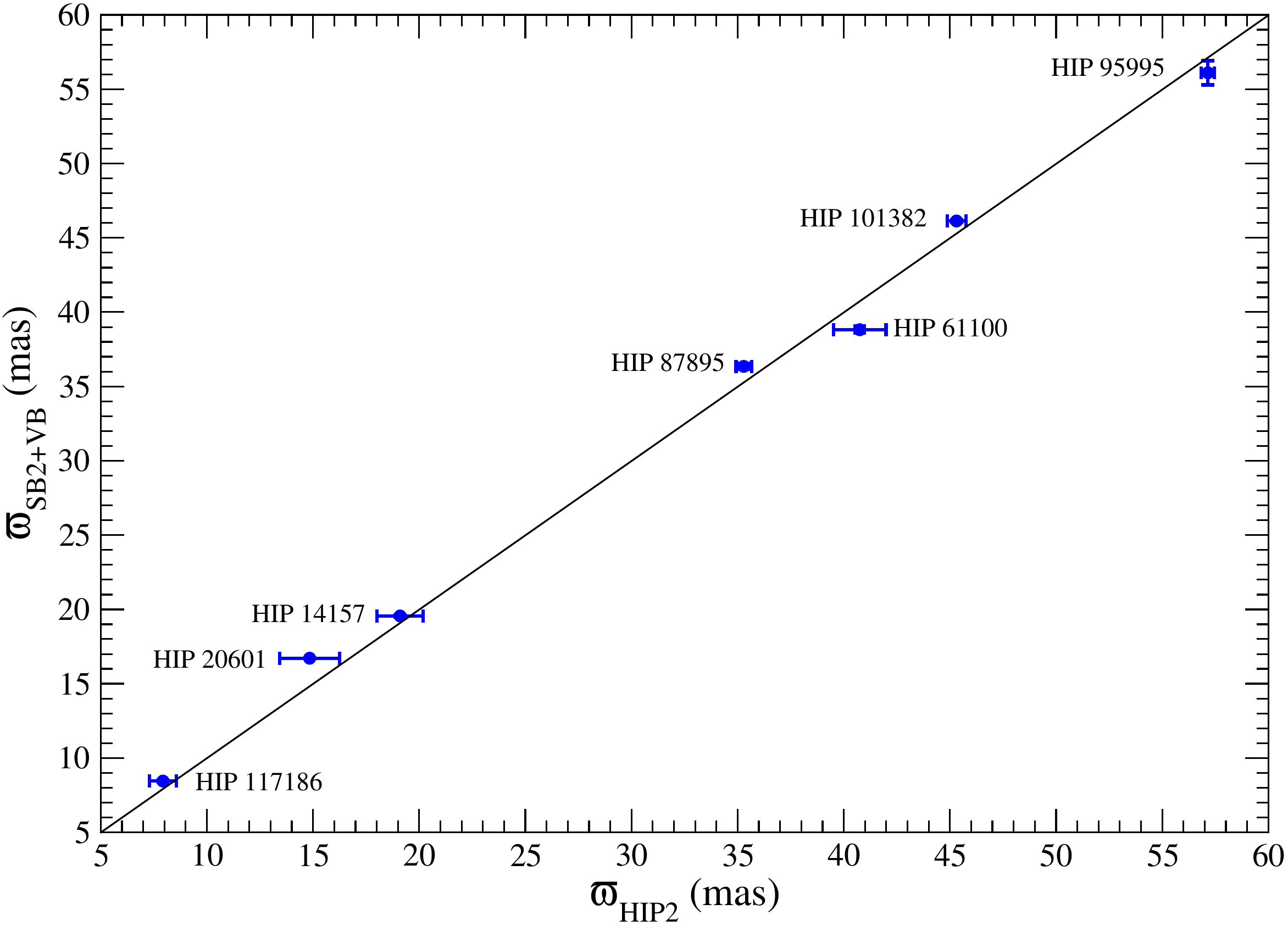}      
%% Note the ABSENCE of the extension .pdf  !
  \caption{Comparison of the corrected Hipparcos 2 parallaxes with the SB2+VB solutions.}
  \label{Halbwachs2:fig5}
\end{figure}

We consider now the binaries for which we have an SB2+VB orbit based on accurate RV measurements.
We count 7 SB2, including 1 from the HERMES programme. Among these binaries, only 3 were observed with PIONIER
because the RV observations are still not completed for the binaries observed by \cite{Boffin17}. The interferometric
measurements of the 4 other binaries are from other sources.

The SB2+VB parallaxes are compared to the corrected Hipparcos 2 parallaxes in Fig.~\ref{Halbwachs2:fig5}.
Since the error bars represent 1 standard deviation, it is obvious that the agreement is rather good. It is also visible
that, HIP 95995 excepted, the uncertainties of the SB2+VB parallaxes are much better than that of Hipparcos 2.
Therefore, they will be usable for the verification of the Gaia parallaxes, at least for the stars brighter than 6th mag.

\section{Conclusions}
%%--------------------

About 80 double stars are observed through three independent programmes in order to improve their SB2 orbital elements.
Thanks to this effort, it will be possible to derive the masses of several
double star components with an accuracy better than 1~\% when the astrometric transits of the Gaia mission will be delivered. 
The reliability of these masses will be verified on the basis
of about a dozen interferometric binaries. For these stars, the parallax is derived from a VB+SB2 orbit, and we
have verified the reliability of the Hipparcos 2 parallaxes. A similar verification will be possible for the 
parallaxes coming from Gaia, at least for the bright stars.

% Optional acknowledgements
% -------------------------

\begin{acknowledgements}

This project was supported by the french INSU-CNRS ``Programme National de Physique Stellaire''
and ``Action Sp\'{e}cifique {\it Gaia}''. PIONIER is funded by the Universit\'e Joseph Fourier (UJF), the Institut de Plan\'etologie et d'Astrophysique de Grenoble (IPAG),
and the Agence Nationale pour la Recherche (ANR-06-BLAN-0421, ANR-10-BLAN-0505, ANR-10-LABX56).
The integrated optics beam combiner is the result of a collaboration between IPAG and CEA-LETI based on CNES R\&T funding. 
We are grateful to the staff of the
Haute--Provence Observatory, and especially to Dr F. Bouchy, Dr H. Le Coroller, Dr M. V\'{e}ron, and the night assistants, for their
kind assistance. We made use of the SIMBAD database, operated at CDS, Strasbourg, France. This research has received funding from the European Community's Seventh Framework Programme (FP7/2007-2013) under grant-agreement numbers 291352 (ERC)

\end{acknowledgements}

%%-----------------------------
%%   Bibliography
%%-----------------------------
%%
%% The reference list should contain all the references cited in the text, ordered alphabetically by surname (with
%% initials following). If there are several references to the same first author, they should be entered according
%% to the following scheme:
%% 1. One author: chronologically
%% 2. Author, one co-author: alphabetically by co-author, then chronologically
%% 3. Author, two or more co-authors: chronologically.
%%
%% Please note that for papers that have more than five authors, only the first three should be given, followed
%% by "et al."
%%
%% The format for references is the one adopted by A&A (see the example below).
%%
%% To set the reference list in the proper A&A format, we encourage you to use BibTEX and the natbib
%% package instead of the standard 'thebibliography' environment.
%%

%% The following lines are required when using BibTEX (strongly encouraged!):
%\bibliographystyle{aa}  % A&A bibliography style file (aa.bst)
%\bibliography{Halbwachs} % your references in file: Yourfile.bib

%
\end{document}